\documentclass[12pt,a4paper]{article}

\usepackage{cite}
\usepackage{latexsym,amssymb,amsmath,amscd,amsthm,amsxtra,graphicx,mathrsfs}

\newcommand{\ba}{\begin{eqnarray}}
\newcommand{\ea}{\end{eqnarray}}
\newcommand{\beq}{\begin{equation}}
\newcommand{\eeq}{\end{equation}}
\newcommand{\bc}{\begin{center}}
\newcommand{\ec}{\end{center}}
\newcommand{\barray}{\begin{array}}
\newcommand{\earray}{\end{array}}

\newcommand{\half}{\hbox{${1\over 2}$}}

\begin{document}

\title{Charged scalar waves from the RN/CFT correspondence}
\author{Xing-Hua Wu, Ran Li\\
{\normalsize \it Department of Physics, Henan Normal University,
 Xin Xiang, 453007, China}}
\date{}
\maketitle

\begin{abstract}
We examine new tests for (non-)extremal 
Reissner-Nordstr\"om/Conformal field
theory correspondences (RN/CFT) in this paper.
The decay rate of the charged scalar wave sourced by an orbiting
star around the black hole is computed and is compared with
the decay rate computed in the corresponding CFT. We find that precise
matches are achieved.


\end{abstract}

\section{Introduction}

As examples of AdS/CFT correspondence \cite{ads-cft},
it is pointed out, e.g.\cite{hmns_cft_Ebh,csz,gg_rn_cft,chz,cs}, 
that the near-horizon dynamics of the charged (RN) black hole 
is equivalent to that of the boundary CFT. For the extremal black hole,
the corresponding CFT is a chiral one \cite{hmns_cft_Ebh,csz,gg_rn_cft,cs}. 
For the near-extremal black hole,
the corresponding CFT has two sectors (left and right)\cite{chz}.

Recently, new checks of AdS/CFT correspondences are suggested
by studying the decay rates of particles excited by some
star orbiting the black holes, see \cite{ps_kerr_cft,bgx_scalar_KN} etc.
In this paper, along this direction, we compute the decay rates of 
the charged scalar particles in two kinds of RN/CFT correspondences
mentioned above \cite{hmns_cft_Ebh,csz,gg_rn_cft,chz,cs}. For these two kinds of
correspondences, on the gravity side, the four dimensional theory
is lifted to a five dimensional one with an extra compact direction 
introduced. So, the
$U(1)$ gauge symmetry in the four dimensional theory becomes an isometry
of the five dimensional RN spacetime metric. It is interesting
to investigate how this geometrization of the 4D gauge symmetry appear
in the computation of the decay rates on the CFT sides.

In the section \ref{sect-nhern}, we consider the near-horizon extremal
RN/Chiral CFT correspondence (NHERN/$\chi$CFT). 
In the section \ref{sect-nhnern}, we 
consider the near-horizon near-extremal RN/CFT correspondence
(NHNERN/CFT). 
For both kinds of correspondences, decay rates computed on the gravity side
match precisely to those computed on the CFT side, respectively. 
We give some remarks on the difference of these two kinds of 
correspondences in the final section
\ref{sect-discuss}.

\section{Near horizon extremal RN/$\chi$CFT}
\label{sect-nhern}

In this section, we will consider the correspondence between
the dynamics of the near-horizon extremal RN black hole and that of the
boundary chiral CFT. Decay rates of a charged scalar wave are computed
on both sides.

\subsection{Decay rate on the gravity side}

The extremal RN black hole is the vacuum solution of the Einstein-Maxwell
theory and is described by the following metric
and the gauge field strength,
\beq \barray{l}
ds^2=-(1-Q/\hat r)^2d\hat t^2+(1-Q/\hat r)^{-2}d\hat r^2 +\hat r^2d\Omega_2^2
\\[10pt]
F=\dfrac{Q}{\hat r^2}d\hat t\wedge d\hat r
\label{Ern-sol} 
\earray\eeq
where $d\Omega_2^2=d\theta^2+\sin^2\theta\, d\varphi^2$ is the round metric
of the two sphere, and one has set the Newton constant $G_N=1$, 
the position of the event horizon is located at $\hat r=M=Q$.
To discuss the near horizon dynamics, taking the limiting process by defining
\cite{gg_rn_cft,csz},
\ba \label{near-horizon-coord}
r={\hat r-Q\over\epsilon}\,,\qquad
t={\epsilon\over Q^2}\hat t
\ea
with $r, t$ fixed when $\epsilon\to 0$. In terms of this new coordinate system,
the metric of black hole and Maxwell field strength becomes,
\ba \label{NHErn-sol} 
ds^2&=&Q^2\left(-r^2dt^2+\dfrac{dr^2}{r^2} +d\Omega_2^2\right)
\\[10pt]
F&=&Qdt\wedge dr
\ea
which shows that the near-horizon geometry is $AdS_2\times S^2$ and the 
isometry group is $SL(2,\mathbb{R})\times SO(3)$.
The gauge potential one-form can be chosen as
\ba
A = -Qr dt
\ea 
by taking a proper gauge.

Let us consider a star carrying electric charge $e_0$ with mass $\mu_0$, 
which is moving around the black hole along an circular orbit
at the near horizon region. For the two Killing vectors, 
$\xi_{(t)}=\partial_t$ and $\xi_{(\varphi)}$, there are two conserved quantities
\cite{mtw_gravitation},
\ba
E_0&=&-g_{\alpha\beta}(u^\alpha+q_0A^\alpha)\xi_{(t)}^\beta
=-Q^2 r_*^2\left(\dot t_*+{q_0\over Qr_*}\right)
\\[5pt]
L_0&=&g_{\alpha\beta}(u^\alpha+q_0A^\alpha)\xi_{(\varphi)}^\beta
=Q^2\sin^2\theta_*\dot\varphi_*
\ea 
where $E_0$ and $L_0$ are energy and angular momentum per unit mass of the star
respectively, $q_0=e_0/\mu_0$, constants $r_*$ and $\theta_*=\pi/2$ 
are radial and angular position of the star respectively.
$L_0$ and $E_0$ are constrained by the on-shell condition,
$-1=Q^2(-r_*^2\dot t_*^2+\sin^2\theta_*\,\dot\varphi_*^2)$. In the
following, without loss of generality, we will set $E_0=0$,  then one have
solution,
\ba
\dot t_*=-{q_0\over Qr_*}\,,\qquad
\dot\varphi_*={L_0\over Q^2}
\ea
with $L_0=Q\sqrt{q_0^2-1}$. We will consider real angular momentum $L_0$, 
i.e., $q_0>1$.

Assume that the coupling of the star with a charged scalar field $\Phi$ is defined by
the following action,
\ba
S_\Phi &=& -\int d^4x\sqrt{-g}\,(\nabla_\mu+ieA_\mu)\Phi^*(\nabla^\mu-ieA^\mu)\Phi
\nonumber\\[5pt]&&
+\,4\pi\lambda\int d\tau\int d^4x\,\delta^4(x-x_*) \Phi^* + {\rm c.c.}
\ea
where $e$ is the electric charge of the scalar field, $\lambda$ is the 
coupling constant, `c.c.' indicates complex conjugate of the previous term. 
The equation of motion reads
\ba
\sqrt{-g}\,(\nabla_\mu-ieA_\mu)(\nabla^\mu-ieA^\mu)\Phi
&=&-4\pi\lambda\int d\tau\, \delta^4(x-x_*)
\nonumber\\[5pt]
&=&-{4\pi\lambda r_*Q\over q_0}\,\delta\left(\varphi+{r_*L_0\over q_0Q}t\right)
\delta(\theta-\pi/2)\delta(r-r_*)
\nonumber
\ea
The source (star) respects the Killing symmetry, 
$\chi=\partial_\varphi - {q_0Q\over r_*L_0}\partial_t$. The variable
separation respects this Killing symmetry is
\ba
\Phi=\sum_{\ell, m} e^{im(\varphi+{r_*L_0\over q_0Q}t)}S_\ell^m(\theta)R_{\ell m}(r)
\equiv \sum_{\ell, m} e^{i(m\varphi-\omega t)}S_\ell^m(\theta)R_{\ell m}(r)
\ea
where, $m\in\mathbb{Z}$, $\omega=-m{r_*L_0\over q_0Q}$. 
The equation of motion separates into radial equations:
\ba \label{ern_req}
r^2R''_{\ell m}(r)  + 2r R'_{\ell m}(r) 
+ \left({\omega^2\over r^2} - {2eQ\omega\over r}
+ e^2Q^2-\mu^2\right)R_{\ell m}(r) = -\tilde\lambda\delta(r-r_*)
\ea
where $\tilde\lambda={2\lambda r_*\over q_0Q}S_\ell^m({\pi\over 2})^*$, 
$\mu^2=\ell(\ell-1)$ ($\ell=0,1,2,\dots$). 
$S_\ell^m(\theta)=\sqrt{(2\ell+1)(\ell-m)!\over 2(\ell+m)!}P_\ell^m(\theta)$ 
is the normalized associated Legendre function of the first kind,
\ba \label{spherical-eigen-eq}
{d^2\over d\theta^2}P_\ell^m(\theta) 
+ \cot\theta\, {d\over d\theta}P_\ell^m(\theta) 
- {m^2\over\sin^2\theta}P_\ell^m(\theta)=-\mu^2P_\ell^m(\theta)
\ea
and the orthonormal condition reads,
\ba
\int d\theta \sin\theta\, S_{\ell'}^m(\theta)^*S_\ell^m(\theta)
=\delta_{\ell\ell'}
\ea 
Two linearly independent solution of the homogeneous radial equation, 
set $\lambda=0$ in (\ref{ern_req}), are given by Whittaker functions,
\ba
R^{(1)}(r) &=& M(-ieQ,h-\half, -2i\omega/r)
\\[5pt]
R^{(2)}(r) &=& W(-ieQ,h-\half, -2i\omega/r)
\ea
where the weight $h$ is defined by
\ba
h=\half +\half\sqrt{1+4\mu^2-4e^2Q^2}
\ea
Near the horizon, 
\ba
R^{(1)}(r\to 0)
&\sim& A e^{-i\omega/r} r^{-ieQ} + B e^{i\omega/r} r^{+ieQ}
\\[5pt]
R^{(2)}(r\to 0)
&\sim& (-2i\omega)^{-ieQ} e^{i\omega/r} r^{+ieQ}
\ea
which shows that $R^{(2)}$ is purely ingoing. Near the boundary, 
\ba
R^{(1)}(r\to \infty) &\sim& (-2i\omega)^h r^{-h}
\\[5pt]
R^{(2)}(r\to \infty) &\sim& C r^{-h} + D r^{h-1}
\ea
We will take $h>1/2$, so
$R^{(1)}$ satisfies purely Neumann boundary condition.
Constants $A, B, C, D$ are defined by
\ba
A=(-2i\omega)^{+ieQ}{\Gamma(2h)\over\Gamma(h+ieQ)}\,,&&
B=(2i\omega)^{-ieQ}e^{i\pi h}{\Gamma(2h)\over\Gamma(h-ieQ)}
\\[5pt]
C=(-2i\omega)^h{\Gamma(1-2h)\over\Gamma(1-h+ieQ)}\,,&&
D=(-2i\omega)^{1-h}{\Gamma(2h-1)\over\Gamma(h+ieQ)}
\ea
Since we want to compute the ingoing flux near the horizon, we
need solution which is purely ingoing at the horizon and satisfies purely
Neumann boundary condition on the boundary. This kind of solution can be
constructed by \cite{ps_kerr_cft}
\ba
R_{\ell m}(r) = \theta(r_*-r)C_2R^{(2)}(r) + \theta(r-r_*)C_1R^{(1)}(r)
\ea
Substituting into the equation of motion, the constants $C_{1,2}$
can be determined to be
\ba
C_1={\tilde\lambda\over 2}{R^{(2)}(r_*)\over W}\,,\qquad
C_2={\tilde\lambda\over 2}{R^{(1)}(r_*)\over W}
\ea
where $W$ is $r$-independent Wronskian,
\ba
W=r^2(R^{(1)}(r)R^{(2)}{}'(r)-R^{(1)}{}'(r)R^{(2)}(r))
=-2i\omega{\Gamma(2h)\over\Gamma(h+ieQ)}
\ea
Now we can compute the Klein-Gordon particle number flux per unit time,
\ba  \label{kg-n-flux}
{\cal F} =-\int d\theta d\varphi\,\sqrt{-g}J^r
\ea
where the current is defined by
\ba \label{scalar-flux-current}
J_\mu = {i\over 8\pi}(\Phi^*(\nabla_\mu-ieA_\mu)\Phi-h.c.)
=-{1\over 4\pi}{\rm Im}(\Phi^*(\nabla_\mu-ieA_\mu)\Phi)
\ea
Then the flux is
\ba
{\cal F} = {Q^4r^2\over 2}\sum_{\ell m}{\rm Im}(R_{\ell m}^*(r)R_{\ell m}'(r))
\ea
Near the boundary, 
\ba \label{horizon-flux}
{\cal F}_{\ell m}(r\to 0) ={Q^4|\omega|\over 2}e^{-\pi eQ}|C_2|^2
\ea
for $m<0$, i.e., $\omega>0$. Near the boundary,
\ba
{\cal F}_{\ell m}(r\to\infty) = 0
\ea
for real $h$.

\subsection{Decay rate on CFT side}

In this subsection, we will compute the decay rate in the CFT on the
boundary dual to the quantum gravity on the near-horizon region
of the extremal RN spacetime. This problem is studied in Refs.
\cite{hmns_cft_Ebh,gg_rn_cft,cs}. Before performing the
computation, let us review some facts here.

The symmetry of the near-horizon-extremal RN black hole includes the isometry
of the metric and the gauge symmetry,
\ba
SL(2,R)\times SO(3)\times U(1)_{\rm em}
\ea
To find out the CFT dual, the four dimensional RN black hole 
is lifted to five dimensions. Especially, the metric becomes,
\ba \label{5d_4d_y}
ds_5^2=ds_{\rm 4D}^2+Q^2(dy+{\cal A})^2
\ea
where $ds_{\rm 4D}^2$ is the 4D RN metric (\ref{NHErn-sol}), the compact
coordinate has period $2\pi$, $y\approx y+2\pi$, ${\cal A}=A/Q$.
In the 5D language, the 4D $U(1)_{\rm em}$ gauge symmetry is geometrized
as the isometry of the $S^1$-fibration of 4D RN black hole.
That is, the 4D gauge symmetry transformation, $A\to A-d\alpha$, appears
in 5D as
\ba  \label{5d-gauge-trans}
{\cal A}\to {\cal A}-{1\over Q}d\alpha\,,\qquad
y\to y+{1\over Q}\alpha\,,\qquad
\Phi\to e^{-ie\alpha}\Phi
\ea
where $\alpha$ is real function depends only on 4D coordinates.

In this 5D setting, the near horizon dynamics will be dual to a chiral 2D
CFT on the boundary: only left-movers are excited and
the right-movers are frozen to the ground state.
For these two sectors, their temperatures are
\ba 
T_L={1\over 2\pi}\,,\qquad T_R=0
\ea 
Specifically, on the
gravity side, non-zero conserved charge of the generator $\partial_t$ 
will indicate the deviation of the extremal state, on the CFT side,
non-zero charge of $SL(2,\mathbb{R})_R$ indicate the excitation of the right-movers
\cite{bhss_superradiance_kerr_cft}.
So we have a correspondence: the time translation $\partial_t$ on gravity side
is mapped to the right-translation $\partial_{\sigma^-}$ on CFT side, or
\ba
t = \sigma^-
\ea
On the other hand, the non-zero conserved charge (electric charge) of
the generator $\partial_y$ on the gravity side corresponds to the non-zero
$SL(2,\mathbb{R})_L$-charge, and $\partial_y$ is mapped to the
left-translation, $\partial_y\sim\partial_{\sigma^+}$, or
\ba
y=-\sigma^+
\ea
where we have used the fact that both $y$ and $\sigma^+$ have period
$2\pi$, so we identify these two coordinate directly, the extra minus sign
is found to be necessary for the matching of the decay rates in
NHERN/$\chi$CFT.

Now, adding an orbiting star into the 4D RN black hole will, on the CFT
side, lead to perturbation of the CFT. Assume that the perturbed action is
\ba
S_{\rm CFT}^{\rm int} 
= \sum_\ell \int d\sigma^+ d\sigma^- J_\ell(\sigma^+,\sigma^-)
{\cal O}_\ell(\sigma^+,\sigma^-)
\ea
The source $J_\ell$ should be determined by the asymptotic behavior of the
scalar field near the boundary, since only scalar field is assumed to be coupled
to the orbiting star. The conformal weight of ${\cal O}$ can also be read off
from the asymptotic behavior of the scalar field, $h_L=h_R=h$.

Now let us look at $J_\ell$. At first, since the star respect the symmetry
$\chi$, so
\ba
J_{\ell}(\sigma^+,\sigma^-) \sim e^{im(\varphi+{r_*L\over q_0Q}t)}
\equiv e^{i(m\varphi-\omega t)}=e^{i(m\varphi-\omega \sigma^-)}
\ea
Secondly, the source $J_\ell$ should have the same gauge behavior as
the charged scalar field. Since a gauge transformation 
is given by (\ref{5d-gauge-trans}), $\Phi\to e^{-ie\alpha}\Phi$, 
we must require the source transform in the same way,
\ba
J_\ell \to e^{-ie\alpha}J_\ell
\ea
and since $y$ is identified as $\sigma^+$ and under gauge transformation,
$y\to y+\alpha/Q$, the dependence of $J_\ell$ on the $y$-coordinate should be
\ba
J_\ell(\sigma^+,\sigma^-) \sim e^{-ieQy}=e^{ieQ\sigma^+}
\ea
In all, we have expansion of the source as follows,
\ba
J_{\ell}(\sigma^+,\sigma^-) =\sum_{m} e^{im\varphi-i\omega \sigma^- +ieQ\sigma^+}J_{\ell m}
\ea
where $J_{\ell m}$ are constants.
To determine $J_{\ell m}$, one can extend the solution $R_{\ell m}$ at the
near horizon region to the whole spacetime \cite{ps_kerr_cft}, 
\ba
R_{\ell m}^{\rm ext}(r)=C_2 R^{(2)}(r)\,,\qquad 0<r<\infty
\ea 
The asymptotic behavior of $R_{\ell m}^{\rm ext}(r)$ is
\ba
R_{\ell m}^{\rm ext}(r\to\infty)\sim C_2 (C r^{-h} + D r^{h-1})
\ea
The coefficient $J_{\ell m}$ can be read off from the Dirichlet mode as,
\ba
J_{\ell m} = C_2 D
\ea
The decay rate of vacuum-to-vacuum per unit time is
\cite{bhss_superradiance_kerr_cft},
\ba \label{dr-nhern-ccft}
{\cal R}&=&2\pi\sum_{\ell m}|J_{\ell m}|^2\int d\sigma^+d\sigma^-\,
e^{i\omega \sigma^- -ieQ\sigma^+}\langle{\cal O}_\ell(\sigma^+,\sigma^-)
{\cal O}_\ell(0,0)\rangle
\nonumber\\[5pt]
&=&2\pi\sum_{\ell m}C_O^2|C_2|^2|D|^2
{(2\pi T_R)^{2h_R-1}\over\Gamma(2h_R)}e^{{\omega\over 2T_R}}
\left|\Gamma\left(h_R+i{\omega\over 2\pi T_R}\right)\right|^2
\nonumber\\[5pt]
&&
\times {(2\pi T_L)^{2h_L-1}\over\Gamma(2h_L)}e^{-{eQ\over 2T_L}}
\left|\Gamma\left(h_L+i{eQ\over 2\pi T_L}\right)\right|^2
\ea
Substituting $T_L={1\over 2\pi}$, $h_L=h_R=h$, and taking the limit,
$T_R\to 0$,
\ba
{\cal R}_{\ell m} 
= C_O^2{2^{5-2h}\pi^2\over Q^{2h+3}(2h-1)^2}{\cal F}_{\ell m}(r\to 0)
\ea
for $m<0$, i.e., $\omega>0$.
This decay rate is precisely equal to that computed
in the gravity side (\ref{horizon-flux}) 
if we take $C_O^2={Q^{2h+3}(2h-1)^2\over 2^{5-2h}\pi^2}$.

\section{Near horizon near-extremal RN/CFT}
\label{sect-nhnern}

There is a correspondence between the dynamics on 
the near-horizon near-extremal RN black hole and a 2d CFT, e.g.\,\cite{chz,cs}.
In this section we will compute the decay rates of a charged scalar wave
from both sides, and compare the results.

\subsection{Decay rate on the gravity side}

Similar to the extremal RN case discussed in the previous section, 
in this subsection, we will consider a charged scalar wave sourced by
a charged star which is orbiting the near-extremal RN black hole.

The non-extremal (general) RN black hole can be described by the metric
and gauge field,
\ba
ds^2&=&-{(\hat r-r_+)(\hat r-r_-)\over \hat r^2}d\hat t^2
+{\hat r^2\over (\hat r-r_+)(\hat r-r_-)}d\hat r^2 +\hat r^2d\Omega_2^2\,,
\\[5pt]
\hat F&=&{Q\over\hat r^2}d\hat t\wedge d\hat r
\ea
Define new coordinates, \cite{chz}
\ba
r={\hat r-Q\over\epsilon}\,,\qquad
t={\hat t\over Q^2}\epsilon\,,\qquad
r_0={\sqrt{2QE}\over\epsilon}
\ea
where $E=Q-M$ is the deviation from the extremality. 
The near-horizon near-extremal limit can be obtained 
by letting $\epsilon\to 0$ and keeping $r, t, r_0$ fixed.
In this new coordinate system, the metric and the gauge field become
\ba
ds^2&=&Q^2\left[-(r^2-r_0^2) dt^2 + {dr^2\over r^2-r_0^2} +d\Omega_2^2\right]
\label{nhnern-met}
\\[5pt]
F&=& Qdt\wedge dr
\ea
where $r=r_0$ is the horizon position. The gauge potential can be chosen as
\ba
A=-Qrdt
\label{nhnern-AF}
\ea
The Hawing temperature of the near-horizon spacetime (\ref{nhnern-met}) 
can be found to be
\ba \label{nhnern-TH}
T_H = {r_0\over 2\pi}
\ea
It is different from the Hawking temperature corresponding to the original
coordinate with hats, $\hat T_H ={\epsilon\over Q^2}T_H$, which
vanishes under the limit $\epsilon\to 0$.

Similar to the extremal case,
corresponding to the two Killing vectors, $\xi_{(t)}=\partial_t$
and $\xi_{(\phi)}$, we define conserved quantities,
\ba
E_0&=&-g_{\alpha\beta}(u^\alpha+q_0A^\alpha)\xi_{(t)}^\beta
=Q^2 (r_*^2- r_0^2)\left(\dot t_*+{q_0r_*\over Q(r_*^2-r_0^2)}\right)
\\[5pt]
L_0&=&g_{\alpha\beta}(u^\alpha+q_0A^\alpha)\xi_{(\varphi)}^\beta
=Q^2\sin^2\theta_*\dot\varphi_*
\ea
where $x_*^{\mu}$ are coordinates of the orbiting star, and we will let
$r=r_*>r_0$ be constant and $\theta=\theta_*=\pi/2$ in the following.
The normalization condition is $-1=Q^2[-(r_*^2-r_0^2)\dot t_*^2+\dot\varphi_*^2]$.
For convenience, take $E_0=0$ and then $\dot t_*=-{q_0r_*\over Q(r_*^2-r_0^2)}$,
$\dot\varphi_*=L_0/Q^2$, or 
\ba
t_*=-{q_0r_*\over Q(r_*^2-r_0^2)}\tau\,,\qquad
\varphi_*={L_0\over Q^2}\tau
\ea
where the integral constants are set to zero. The equation of motion
of the scalar field is
\ba
\sqrt{-g}\,(\nabla_\mu-ieA_\mu)(\nabla^\mu-ieA^\mu)\Phi
=-{4\pi\lambda\over |\dot t_*|}\,
\delta\left(\varphi-{\dot\varphi_*\over\dot t_*}t\right)
\delta(\theta-\pi/2)\delta(r-r_0)
\ea
The source (star) respects the Killing symmetry, 
$\chi=\partial_\varphi+{\dot t_*\over\dot\varphi_*}\partial_t$. The variable
separation respects this Killing symmetry is
\ba
\Phi=\sum_{\ell, m} e^{im(\varphi-{\dot\varphi_*\over\dot t_*}t)}
S_\ell^m(\theta)R_{\ell m}(r)
\equiv \sum_{\ell, m} e^{i(m\varphi-\omega t)}S_\ell^m(\theta)R_{\ell m}(r)
\ea
where $\omega=m{\dot\varphi_*/\dot t_*}$, $m\in\mathbb{Z}$.
In the following we will take $m>0$, i.e., $\omega<0$, 
in the concrete calculation.
The equation of motion separates into radial equations:
\ba \label{nern_req}
((r^2-r_0^2)R'_{\ell m})'
+ \left[{(\omega-eQr)^2\over r^2-r_0^2}-\mu^2\right]R_{\ell m}
= -\tilde\lambda\delta(r-r_*)
\ea
where $\mu^2=\ell(\ell-1)$ ($\ell=0,1,2,\dots$), $R_{\ell m}'=\partial_rR_{\ell m}$,
$\tilde\lambda={2\lambda\over|\dot t_*|Q^2}S_\ell^m({\pi\over 2})^*$.
$S_\ell^m(\theta)$ is defined in the previous section.
The two linearly independent solutions of the radial homogeneous equation 
(when $\lambda=0$) are the following hyper-geometric functions,
\ba  \label{hypergeometric-1}
R_{\ell m}^{(1)}=(r+r_0)^{-i\alpha}(r-r_0)^{i\beta}\,
{}_2F_1\left(1-h-i\alpha+i\beta, h-i\alpha+i\beta, 1-2i\alpha,
{r+r_0\over 2r_0}\right)
\\[5pt]  \label{hypergeometric-2}
R_{\ell m}^{(2)}=(r+r_0)^{i\alpha}(r-r_0)^{i\beta}\,
{}_2F_1\left(1-h+i\alpha+i\beta, h+i\alpha+i\beta, 1+2i\alpha,
{r+r_0\over 2r_0}\right)
\ea 
where the constants are defined by
\ba \label{parameter-def}
\alpha=\hbox{${1\over 2}|eQ+{\omega\over r_0}|$}\,,\quad
\beta=\hbox{${1\over 2}(eQ-{\omega\over r_0})$}\,,\quad
\hbox{$h={1\over 2}+{1\over 2}\sqrt{1-4e^2Q^2+4\mu^2}$}
\ea
The asymptotic behaviors near the boundary of $R^{(1,2)}$ are
\ba
R_{\ell m}^{(1)}(r\to\infty)\sim A_\infty r^{h-1} + B_\infty r^{-h}
\\[5pt]
R_{\ell m}^{(2)}(r\to\infty)\sim C_\infty r^{h-1} + D_\infty r^{-h}
\ea
where the coefficient $A_\infty$ is defined by
\ba
A_\infty=\frac{\Gamma (1-2 i \alpha ) \Gamma (2 h -1) 
(-2r_0)^{-i \alpha +i \beta -h +1} }
{\Gamma (h-i\alpha -i\beta ) \Gamma (h -i \alpha +i \beta)}
\ea
and the other coefficients are related to $A_\infty$: $B_\infty=A_\infty(h\to 1-h)$,
$C_\infty=A_\infty(\alpha\to -\alpha)$ and 
$D_\infty=A_\infty(h\to 1-h, \alpha\to -\alpha)$.
The near horizon behaviors are
\ba
R_{\ell m}^{(1)}(r\to r_0)\sim A_0 (r-r_0)^{i\beta} + B_0 (r-r_0)^{-i\beta}
\\[5pt]
R_{\ell m}^{(2)}(r\to r_0)\sim C_0 (r-r_0)^{i\beta} + D_0 (r-r_0)^{-i\beta}
\ea
where the coefficient $A_0$ is defined by
\ba
A_0={i \pi  (2r_0)^{-i \alpha }{\rm csch}(2 \pi  \beta ) \Gamma (1-2 i \alpha )  
\over\Gamma (2 i \beta +1) \Gamma (1-h-i \alpha -i\beta) 
\Gamma (h-i \alpha -i\beta)}
\ea
and the other coefficients are related to $A_0$:
$B_0=e^{2\pi\beta}(2r_0)^{2i\beta}A_0(\beta\to -\beta)$,
$C_0=A_0(\alpha\to -\alpha)$ and $D_0 = B_0(\alpha\to -\alpha)$.
From these results, one can see that the mode which is purely ingoing 
near horizon (near region) can be constructed as
\ba
R_{\ell m}^{(n)} = D_0 R_{\ell m}^{(1)} - B_0 R_{\ell m}^{(2)}
\ea
and the mode which is purely Neumann at infinity (far region) is
\ba
R_{\ell m}^{(f)} = C_\infty R_{\ell m}^{(1)} - A_\infty R_{\ell m}^{(2)}
\ea
We have
\ba
R_{\ell m}^{(n)}(r\to r_0)
&\sim& {\alpha\over\beta} (2r_0)^{2i\beta}e^{2\pi\beta}(r-r_0)^{i\beta}
\\[5pt]
R_{\ell m}^{(f)}(r\to\infty)
&\sim& {2i\alpha\over 1-2h}(2r_0)^{1+2i\beta}e^{2\pi\beta}r^{-h}
\ea
Now the solution of the inhomogeneous equation (\ref{ern_req}) is constructed
by the Green's method as
\ba
R_{\ell m} = \theta(r_*-r)C_{(n)}R_{\ell m}^{(n)}
+\theta(r-r_*)C_{(f)}R_{\ell m}^{(f)}
\ea
where 
\ba
C_{(n)}={\tilde\lambda\over 2}{R^{(f)}(r_*)\over W}\,,\qquad
C_{(f)}={\tilde\lambda\over 2}{R^{(n)}(r_*)\over W}
\ea
where $W$ is $r$-independent Wronskian,
\ba
W&=&(r^2-r_0^2)(R^{(n)}(r)R^{(f)}{}'(r)-R^{(n)}{}'(r)R^{(f)}(r))
\nonumber\\[5pt]
&=&2\pi i\alpha^2(2r_0)^{2-h+5i\beta}e^{2\pi\beta}
{(1+\coth\pi\beta)^2\tanh(\pi\beta)\Gamma(2h-1)
\over\Gamma(1-2i\beta)\Gamma(h-i\alpha+i\beta)\Gamma(h+i\alpha+i\beta)}
\ea
The Klein-Gordon particle number flux per unit time is still given by the equation
(\ref{kg-n-flux}) and for the mode with quantum numbers $\ell, m$,
\ba
{\cal F}_{\ell m} = {Q^2\over 2}(r^2-r_0^2)\,{\rm Im}(R_{\ell m}^*(r)R_{\ell m}'(r))
\ea
At infinity, simple calculation shows that ${\cal F}_{\ell m}(r\to\infty)=0$.
Near the horizon,
\ba
{\cal F}_{\ell m}(r\to r_0)={\alpha^2\over\beta}Q^2r_0 e^{4\pi\beta}|C_{(n)}|^2
\ea
Since the spacetime has non-zero temperature (\ref{nhnern-TH}), the decay rate
of the scalar particle into the horizon is
\ba
{\cal R}_{\ell m}={1\over e^{\omega+e\Phi_H\over T_H}-1}{\cal F}_{\ell m}(r\to r_0)
\ea
where $\Phi_H=A_t(r=r_0)=-Qr_0$ is the gauge potential at the horizon (chemical
potential). Using the definition of $\beta$ in (\ref{parameter-def}), we have
\ba \label{decay-rate-gravity-side}
{\cal R}_{\ell m}={1\over 1-e^{-4\pi\beta}}{\alpha^2\over\beta}Q^2r_0|C_{(n)}|^2
\ea
We will see that this Planck factor is important for the matching of the decay
rates in NHNERN/CFT correspondence.

\subsection{Decay rate on the CFT side}

Now turn to the computation from the CFT viewpoint. The analysis of the dictionary
in the near-horizon near-extremal RN/CFT correspondence is very similar
to the one in the near-horizon extremal RN/$\chi$CFT correspondence, which is
discussed in the previous section. For details, one can refer to the original
references, \cite{chz,cs} etal. In the following we will only sketch points
necessary in the computation.

Similar to the extremal case discussed in the previous section,
we have the following matching of coordinates,
\ba
t\leftrightarrow\sigma^-\,,\qquad y\leftrightarrow -\sigma^+
\ea
Corresponding to this identification, one can derive the left and right
temperatures of the boundary CFT as \cite{chz},
\ba
T_L={1\over 2\pi}\,,\qquad T_R={r_0\over 2\pi}
\ea
Furthermore, the conformal weights of the operator corresponding to the
scalar field $\Phi$ is $(h_L,h_R)=(h,h)$. 
Similar to the extremal case, the source $J_\ell$ depends on the compact
coordinate through 
$J_{\ell}(\sigma^+,\sigma^-) 
=\sum_{m} e^{im\varphi-i\omega \sigma^- +ieQ\sigma^+}J_{\ell m}$.
Now using the expression
(\ref{dr-nhern-ccft}) for the decay rate, for the mode with quantum numbers
$\ell, m$,
\ba
{\cal R}_{\ell m}^{(\rm CFT)}
=2\pi e^{-2\pi\beta} \left({r_0\over Q}\right)^{2h-1}
{|\Gamma(h-i{\omega/r_0})|^2|\Gamma(h+ieQ)|^2
\over\Gamma(2h)^2}
C_O^2|J_{\ell m}|^2
\ea
The constant $J_{\ell m}$ is determined by the Dirichlet term of 
$R_{\ell m}^{(n)}$ near infinity. That is, it is the coefficient of $r^{h-1}$ in
$c_{(n)}R_{\ell m}^{(n)}$ when $r\to\infty$,
\ba
J_{\ell m} = {4\pi\alpha(2r_0)^{1-h+3i\beta}\over 1-e^{-4\pi\beta}}
{\Gamma(2h-1)
\over\Gamma(1-2i\beta)\Gamma(h-i\alpha+i\beta)\Gamma(h+i\alpha+i\beta)}
C_{(n)}
\ea
Substituting into the decay rate, we have
\ba
{\cal R}_{\ell m}^{(\rm CFT)}
={1\over 1-e^{-4\pi\beta}}{\alpha^2\over\beta}
{2^{5-2h}\pi^2r_0Q^{1-2h}\over (2h-1)^2}C_O^2|C_{(n)}|^2
\ea
If we choose $C_O^2={Q^{2h+1}(2h-1)^2\over 2^{5-2h}\pi^2}$, this decay rate
is completely the same as that computed in the gravity side 
(\ref{decay-rate-gravity-side}).

\section{Discussions}
\label{sect-discuss}

If we compare the computations of the decay rates in the previous
two sections, one will find that if the extremal limit, $E=M-Q\to 0$,
is taken before the near horizon limit, $\epsilon\to 0$, the near-extremal
case reduced to the extremal one ($r_0=0$). However, if the near horizon limit
is taken before the extremal limit, we will obtain the near-extremal
case ($r_0\neq 0$). It should be pointed out that one could not obtain 
the results of the extremal case by taking the limit $r_0\to 0$ on the 
results of the near-extremal case: $r_0=0$ is the essential singularity
of the hypergeometric functions (\ref{hypergeometric-1})
and (\ref{hypergeometric-2}).

\bigskip\bigskip

\noindent
{\bf Acknowledgments.}
R. L. is supported by National Natural Science Foundation of China
(Grant No. 11205048). 
R. L. is also supported by the Foundation for Young Key Teacher of 
Henan Normal University.


\begin{thebibliography}{99}

\bibitem{ads-cft}
J. M. Maldacena, 
The large N limit of superconformal field theories and supergravity, 
Int. J. Theor. Phys. {\bf 38} (1999) 1113.

S.S. Gubser, I.R. Klebanov, A.M. Polyakov,
Gauge Theory Correlators from Non-Critical String Theory,
Phys. Lett. {\bf B428} (1998) 105.

E. Witten,
Anti De Sitter Space And Holography,
Adv. Theor. Math. Phys. {\bf 2} (1998) 253.

\bibitem{hmns_cft_Ebh}
T. Hartman, K. Murata, T. Nishioka and A. Strominger,
CFT duals for extreme black holes,
JHEP 0904 (2009) 019.

\bibitem{gg_rn_cft}
A. Ghodsi, M.R. Garousi,
The RN/CFT Correspondence,
Phys. Lett. {\bf B687} (2010) 79.

\bibitem{csz}
Chiang-Mei Chen, Jia-Rui Sun, Shou-Jyun Zou,
The RN/CFT Correspondence Revisited,
JHEP 1001 (2010) 057.

\bibitem{chz}
Chiang-Mei Chen, Ying-Ming Huang, Shou-Jyun Zou,
Holographic Duals of Near-extremal Reissner-Nordstrom Black Holes,
JHEP {\bf 1003} (2010) 123.

\bibitem{cs}
Chiang-Mei Chen, Jia-Rui Sun, 
Holographic Dual of the Reissner-Nordstr\"om Black Hole,
J. Phys. Conf. Ser. {\bf 330} (2011) 012009.


\bibitem{ps_kerr_cft}
A. P. Porfyriadis, A. Strominger,
Gravity Waves from Kerr/CFT,
Phys. Rev. {\bf D90} (2014) 044038.

\bibitem{bgx_scalar_KN}
N. Bai, Y.-H. Gao, X.-B. Xu,
On neutral scalar radiation by a massive orbiting star in extremal Kerr-Newman black hole,
Fortsch. Phys. {\bf 63} (2015) 323, arXiv:1407.0089.

\bibitem{mtw_gravitation}
C.W. Misner, K.S. Thorne, J.A. Wheeler,
Gravitation, \S 33.5, Freeman, 1973.

\bibitem{bhss_superradiance_kerr_cft}
I. Bredberg, T. Hartman, W. Song, A. Strominger,
Black Hole Superradiance From Kerr/CFT,
JHEP 1004 (2010) 019.




\end{thebibliography}
\end{document}